\documentclass[10pt, conference]{IEEEtran}
\usepackage[dvips]{graphicx}
\usepackage{amsmath}
\usepackage{subfigure}
\usepackage{epsfig, psfrag, amsmath, amssymb, amsfonts, latexsym, balance,indentfirst,}
\usepackage[numbers,sort&compress]{natbib}
\usepackage{algorithm, algorithmic}
\usepackage{multirow}
\usepackage{amsthm}
\usepackage{xcolor}
\usepackage{extarrows}
\usepackage[small]{caption2}
\allowdisplaybreaks
\newlength{\figwidth}
\ifCLASSOPTIONonecolumn
\usepackage{geometry}
\usepackage{lipsum}
\else
\fi
\begin{document}

%%%%%%%%%%%%%%%%%%%%%%%%%%%%%%%%%%%%%%%%%%%%%%%%%%%%%%%%%%%%%%%%%%%%%%%%%%%%%%%%%%%%%%%%%%%%%%%%%%%%%%%%%%%%
\title{Functional Gaussian Distribution Modelling of Mobility Prediction Accuracy for Wireless Users}

\author{
	\IEEEauthorblockN{Lu Liu, Wuyang Zhou, Sihai Zhang}
\IEEEauthorblockA{Key Laboratory of Wireless-Optical Communications, \\University of Science and Technology of China\\  Hefei, Anhui, China 230027\\
Email: liuluzz@mail.ustc.edu.cn,  \{wyzhou, shzhang\}@ustc.edu.cn}

\and

	\IEEEauthorblockN{Wei Cai}
\IEEEauthorblockA{School of Information and Science and Technology, \\University of Science and Technology of China\\  Hefei, Anhui, China 230027\\
	Email: caiwei@ustc.edu.cn}
}

\maketitle
\thispagestyle{empty}
\begin{abstract}	
Mobility entropy is proposed to measure predictability of human movements, based on which, the upper and lower bound of prediction accuracy is deduced, but corresponding mathematical expressions of prediction accuracy keeps yet open.
In this work, we try to analyze and model prediction accuracy in terms of entropy based on the 2-order Markov chain model empirical results on a large scale CDR data set, which demonstrates the observation that users with the same level of entropy achieve different levels of accuracy\cite{Empirical}.
After dividing entropy into intervals, we fit the probability density distributions of accuracy in each entropy interval with Gaussian distribution and then we estimate the corresponding mean and standard deviation of these distributions.
After observing that the parameters vary with increasing entropy, we then model the relationship between parameters and entropy using least squares method. The mean can be modelled as a linear function, while the standard deviation can be modelled as a Gaussian distribution.
Thus based on the above analysis, the probability density function of accuracy given entropy can be expressed by functional Gaussian distribution.
The insights from our work is the first step to model the correlation prediction accuracy and predictability entropy, thus shed light on the further work in this direction.

\end{abstract}


\section{Introduction}
% The importance of mobility prediction

Human mobility prediction plays a critical role in many fields, such as urban planning\cite{urban planning}, emergency management\cite{emergency management}, transportation engineering\cite{transportation engineering}, public health\cite{public health} and location-based services\cite{location-based services}.
Various methods have been proposed to predict human mobility, such as Markov chain models \cite{markov chains}, artificial neural networks \cite{neural network,neural network2}, Bayesian networks \cite{bayesian,HMM}, and so on.
It has been now recognized that prediction accuracy relies on prediction models and data sets.  \cite{markov chains} implements mobility prediction based on $n$-order Markov-based model and figures out that the 2-order Markov-based model has the highest accuracy, while \cite{approaching} claims that 1-order Markov performs better.
% Works related to predictability

After Gonz\'alez et al. \cite{understanding} demonstrate that human movements show a high degree of spatial and temporal regularity, Song et al. \cite{predictability} propose an approach to measure the ability of predicting the whereabouts and mobility of individuals.
They adopt mobility entropy to characterize the predictability of human mobility, where higher entropy indicates less predictable.
Following this work, several research works have used this metric to explore the limits of predictability in human dynamics\cite{patterns,approaching,friendship}.

% Our Work

According to the theoretical entropy, though, there are one emerging important question: Can prediction accuracy be described using the mobility entropy?
In other words, if someone's mobility entropy is known, can we tell its prediction accuracy using mathematical formula without after actual prediction, even by just a specific prediction model?
This paper is trying to answer this question.
To be specific, \cite{predictability} give a upper bound of prediction accuracy, and this work tries to build a mathematical model to describe the relationship between the achieved accuracy and predictability based on a real data set, even using one certain prediction model.
Note that our previous work\cite{Empirical} has already provided the users' predictability and their accuracy after predictability calculation and 2-order Markov prediction.

%Based on a CDR data set, we make prediction based on 2-order Markov chain model and obtain the distribution of the prediction accuracy. In the several literatures, 2-order Markov chain model have good performance. So it is convincing to use it to analyze.

In order to analyse accuracies of users with the same level of entropy, entropy value are divided into multiple intervals. We analyze the probability density distributions of prediction accuracy corresponding to the intervals and find that the distributions approximate Gaussian distributions. Then, we fit the distribution of accuracy with Gaussian distribution and obtain the corresponding parameters, mean, $\mu$, and standard deviation, $\sigma$, which vary with the increasing entropy. The relationship between the mean and entropy can be modelled as a linear function. As for the standard deviation, the relationship is not so simple to describe. We try three distributions and choose a Gaussian distribution based on MSE(mean square error) comparison at last. Finally, we draw a functional Gaussian distribution to describe the distributions of accuracy given entropy.

The contributions of this paper are as following:

\begin{itemize}
	
\item We model the prediction accuracy based on predictability entropy. Different from \cite{predictability}, which deduces the upper and lower bound of prediction accuracy based on entropy, this work is pioneering to describe the correlation between accuracy and entropy in the form of probability density distribution. Although this work is just based on one real world CDR data set and 2-order Markov model, it sheds light on future work in this direction.

\item We provide an inference approach to model the correlations and estimate the parameters. We also conclude that correlations can be modelled as a linear function and Gaussian distribution with $\mu=-0.1726s+0.9845$ and $\sigma=0.09415exp(-(\frac{s-2.548}{1.96})^2)$, where $s$ is discrete entropy value. Following work may take this way to investigate more data sets and prediction models to obtain more general conclusions.

\end{itemize}

%The rest of paper is organized as following. Section \ref{sec2} presents our data set and processing methodology. Section \ref{sec3} illustrates the measurement of predictability and prediction model, showing entropy and prediction accuracy results of users. Section \ref{sec4} present the detail that how to model the accuracy based on entropy. Section \ref{sec5} makes discussions about our work. Finally, Section \ref{sec6} concludes the paper.

\section{Data Sets}
\label{sec2}

In this section, the details of the real mobile CDR data set used in this work will be introduced, which is provided by a Chinese telecom operator.
This data set starts from Tuesday, July 1, 2014, and ends on Wednesday, Dec. 31, 2014, which possesses 194,336 anonymous mobile phone subscribers registering in one Chinese city.
Totally 270,932,374 records provide the spatial and temporal information of telecom users. The record format is shown as below:

(SERVICE\underline{ }NBR, CALL\underline{ }TYPE, OPPOSITE\underline{ }NO, TOLLTYPE\underline{ }ID, ROAM\underline{ }TYPE, START\underline{ }TIME, END\underline{ }TIME, DURATION, CITY\underline{ }ID, ROAM\underline{ }CITY\underline{ }ID, OPPCITY\underline{ }ID and  LAC\underline{ }ID)

LAC\underline{ }ID(Location Area Code ID) is adopted as the user's location, and 453,752 locations are included in the data set. In order to extract the trajectories of users, we take the spatial and temporal features and remove the others. Then, the four features are left: SERVICE\underline{ }NBR, START\underline{ }TIME, ROAM\underline{ }CITY\underline{ }ID, LAC\underline{ }ID, for predictability calculation and next place prediction.

In order to make sure that every prediction object has enough spatial and temporal information, we define the active day concept as a user has at least one location update on that day and remove the users whose active days are less than 150. After the processing, 142,288 users with total 247,780,761 records are left.

\section{Predictability Entropy and Prediction Accuracy}
\label{sec3}
In this section, we describe the basic data of the work. First, we introduce the predictability measurement, entropy, and show the predictability measurement results of the users in the CDR data. Then, we show the prediction model and correlation between accuracy and entropy.
\subsection{Entropy Calculation}
Entropy is a basic quantity for measuring the degree of predictability characterizing a time series in spatial-temporal space\cite{predictability}. The definitions of the three entropy measures for each individual's mobility pattern are briefly described below and more details are shown in the original literatures mentioned above.

\emph{i) Random entropy}: This entropy focuses on the number of unique locations that a person visits, assuming that each location is visited with equal probability by one user. Thus, it is defined as $S^{\text{rand}}={\rm{log}_{2}}N$, where $N$ is the number of unique locations to which a user has traveled.

\emph{ii) Temporal-uncorrelated entropy}: This entropy takes the visiting frequency of locations into consideration. It is defined as $S^{\text{unc}}=-\sum_{i}p(l_{i}){\rm{log}_2}p(l_{i})$, where $p(l_{i})$ is the probability that location $l_{i}$ was visited by the user.

\emph{iii) Real entropy}: This entropy captures the full spatial-temporal order presented in a user's mobility pattern. Suppose the location sequence of a person's trajectory is $Traj=l_{1}\rightarrow l_{2}\rightarrow l_{3}\rightarrow\cdots$. The real entropy can be calculated as

\begin{equation}\label{eq1}
S^{\text{real}}=-\sum_{Traj^{\prime}_{i}}p(Traj^{\prime}_{i}){\rm{log}_{2}}p(Traj^{\prime}_{i}),
\end{equation}
where $Traj^{\prime}_{i}$ is a subsequence of $Traj$ and $p(Traj^{\prime}_{i})$ is the probability of $Traj^{\prime}_{i}$ appearing in $Traj$.

\begin{figure}	
\centerline{\includegraphics[width=0.4\textwidth]{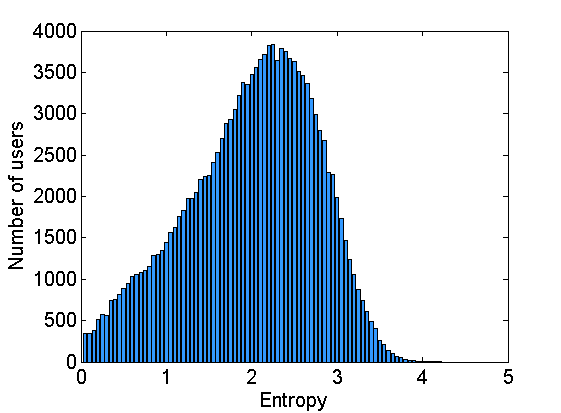}}
\caption{The distribution of entropy in the CDR data.}
\label{Entropy}
\end{figure}

As real entropy incorporates a person's spatial and temporal features, we only consider real entropy of users in this paper. Based on the previous work, we find that users with the same entropy have different prediction accuracies. In order to figure out the characteristic of the accuracy of users with the same level of entropy, we divide the range of entropy into intervals, length of $0.05$. There are 84 entropy intervals in total. The number of users in each interval is shown in the Fig. \ref{Entropy}.

\subsection{Next Place Prediction}

2-order Markov chain model is an effective prediction approach to make next place prediction. Thus we set it as prediction model. It assumes that an individual's next place is related to the previous 2 visited locations and the trajectory of an individual can be modeled as a 2-order Markov chain. Based on the CDR data, transition probability can be calculated.

\begin{equation}
\begin{aligned}
\label{eq4}
&P(X^{t+1}=l_{t+1}|X^{t}=l_{t},...,X^{1}=l_{1})\quad\quad\quad\quad\\
&=P(X^{t+1}=l_{t+1}|X^{t}=l_{t},X^{t-1}=l_{t-1}).
\end{aligned}
\end{equation}

\noindent where $X^{t}$ is a variable representing the location for an individual at step $t$.

According to the transition matrix $P$, we can make next place prediction. The location corresponding to the maximum transition probability is chosen as the next place.

\begin{figure}	
\centerline{\includegraphics[width=0.4\textwidth]{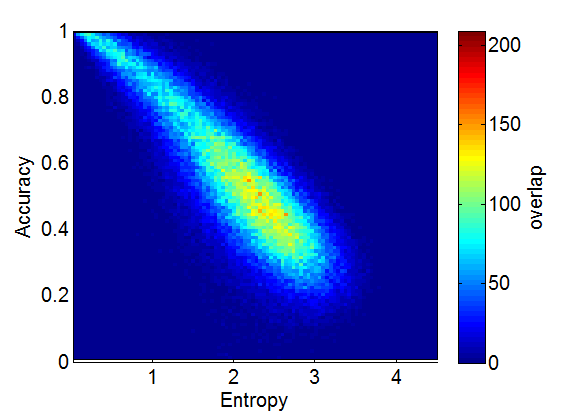}}
\caption{The correlation between predictability entropy and prediction accuracy \cite{Empirical}. }
\label{Acc-S}
\end{figure}

Based on the entropy and prediction accuracy of every user, we can get the relationship between predictability entropy and prediction accuracy as shown in Fig. \ref{Acc-S}. Obviously, users with the same entropy achieve different levels of accuracy. In the next section, we try to model the accuracy corresponding to different levels of entropy.

\section{Prediction Accuracy Modelling}
\label{sec4}
In this section, we model the prediction accuracy based on the predictability entropy and the prediction result. The probability density distribution of accuracy can be fitted with Gaussian distribution, and we get the parameters, mean and standard deviation, respectively. The parameters vary with increasing entropy and they can be estimated by least squares method. After estimation, we draw a formula to describe the the distribution of accuracy given entropy.

\subsection{Accuracy Modelling Based on Entropy}
Observing the correlation between accuracy and entropy, we are surprised that users with the same entropy achieve various prediction accuracies. Then, we decide to research the distribution of achieved accuracy of the users with the same entropy. In Section \ref{sec3}, the range of entropy are divided into 84 intervals, and we analyze the results of accuracy in the same entropy intervals in this section.

In Fig. \ref{Acc-S}, we observe that the distribution of accuracy in every entropy interval is similar to unimodel distribution and the distribution shows a degree of symmetry. Accordingly, it is reasonable to fit the distribution of accuracy given entropy with Gaussian distribution.

A general formal of Gaussian distribution is shown below:

\begin{equation}
\begin{aligned}
\label{normal}
&P(x|s)=\frac{1}{\sqrt{2\pi}\sigma}exp(-\frac{(x-\mu)^2}{2\sigma^2}), -\infty<x<\infty.\\
\end{aligned}
\end{equation}
where $x$ can be used to represent accuracy, $s$ is denoted as entropy, $\mu$ can represent the mean of accuracy, and $\sigma$ can equal the standard deviation of accuracy.

In Fig. \ref{Acc-S}, we can see that users with higher entropy achieve lower accuracy. Thus, $\mu$ decreases with increasing entropy. The size of accuracy range in the different entropy intervals also varies, and then $\sigma$ is also a variable changing with entropy. We fit the probability density distribution of accuracy and make parameter estimation in the following subsections.

\subsection{Fitting Accuracy of Entropy Interval}
\begin{figure*}[!htb]
\centering
\subfigure[]{
\begin{minipage}[b]{0.3\textwidth}
\includegraphics[width=1\textwidth]{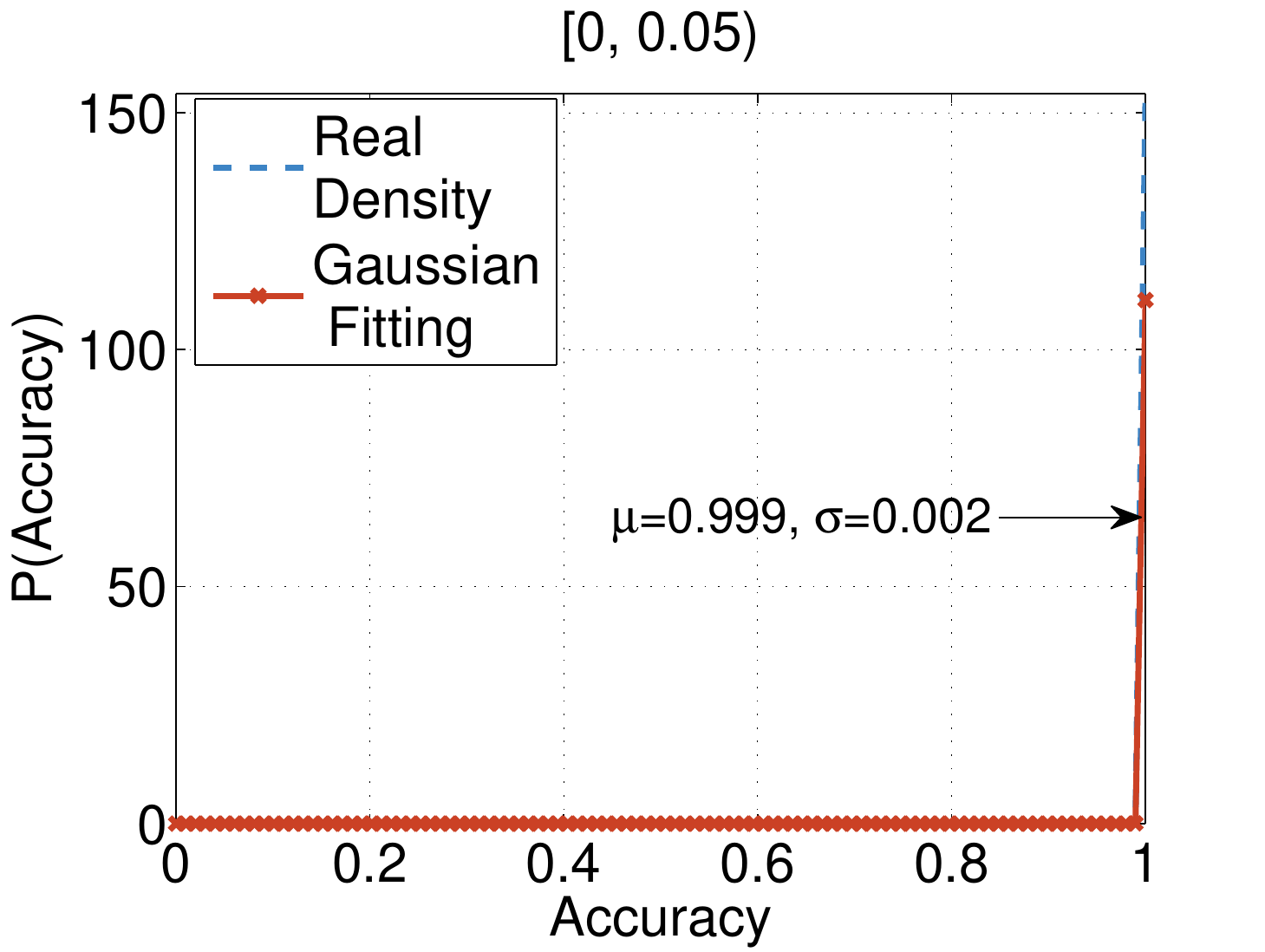}
\end{minipage}
}
\subfigure[]{
\begin{minipage}[b]{0.3\textwidth}
\includegraphics[width=1\textwidth]{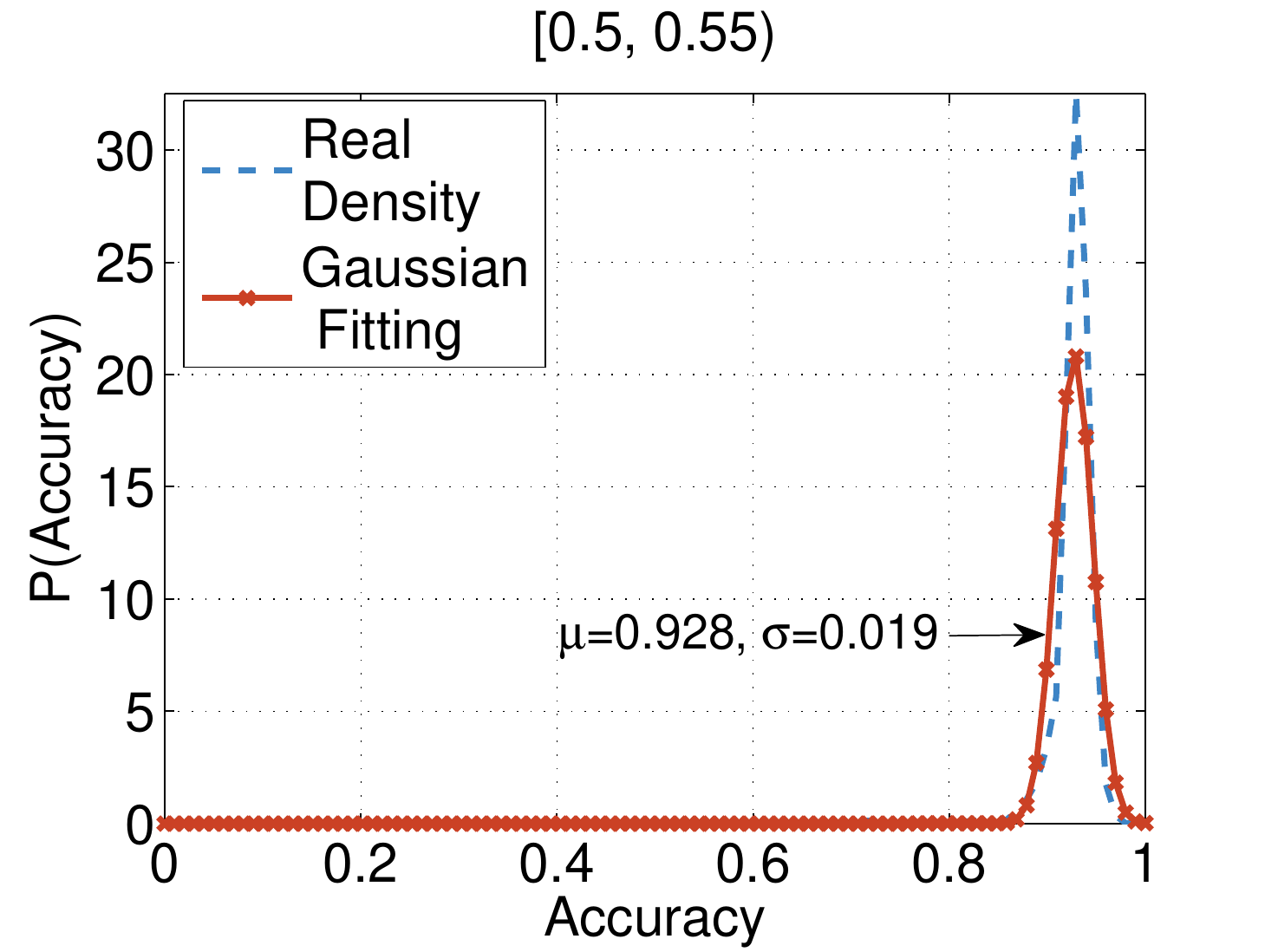}
\end{minipage}
}
\subfigure[]{
\begin{minipage}[b]{0.3\textwidth}
\includegraphics[width=1\textwidth]{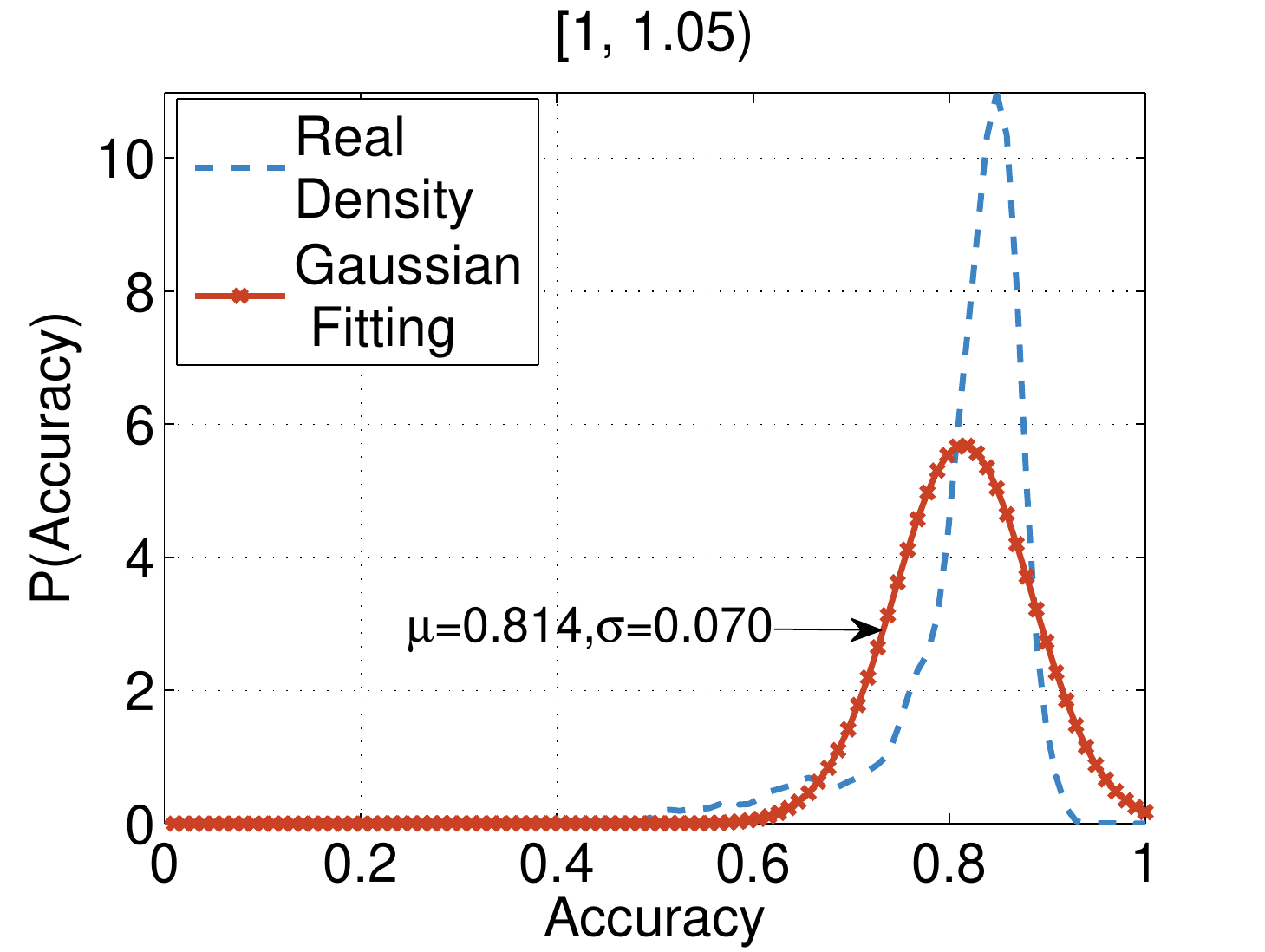}
\end{minipage}
}
\subfigure[]{
\begin{minipage}[b]{0.3\textwidth}
\includegraphics[width=1\textwidth]{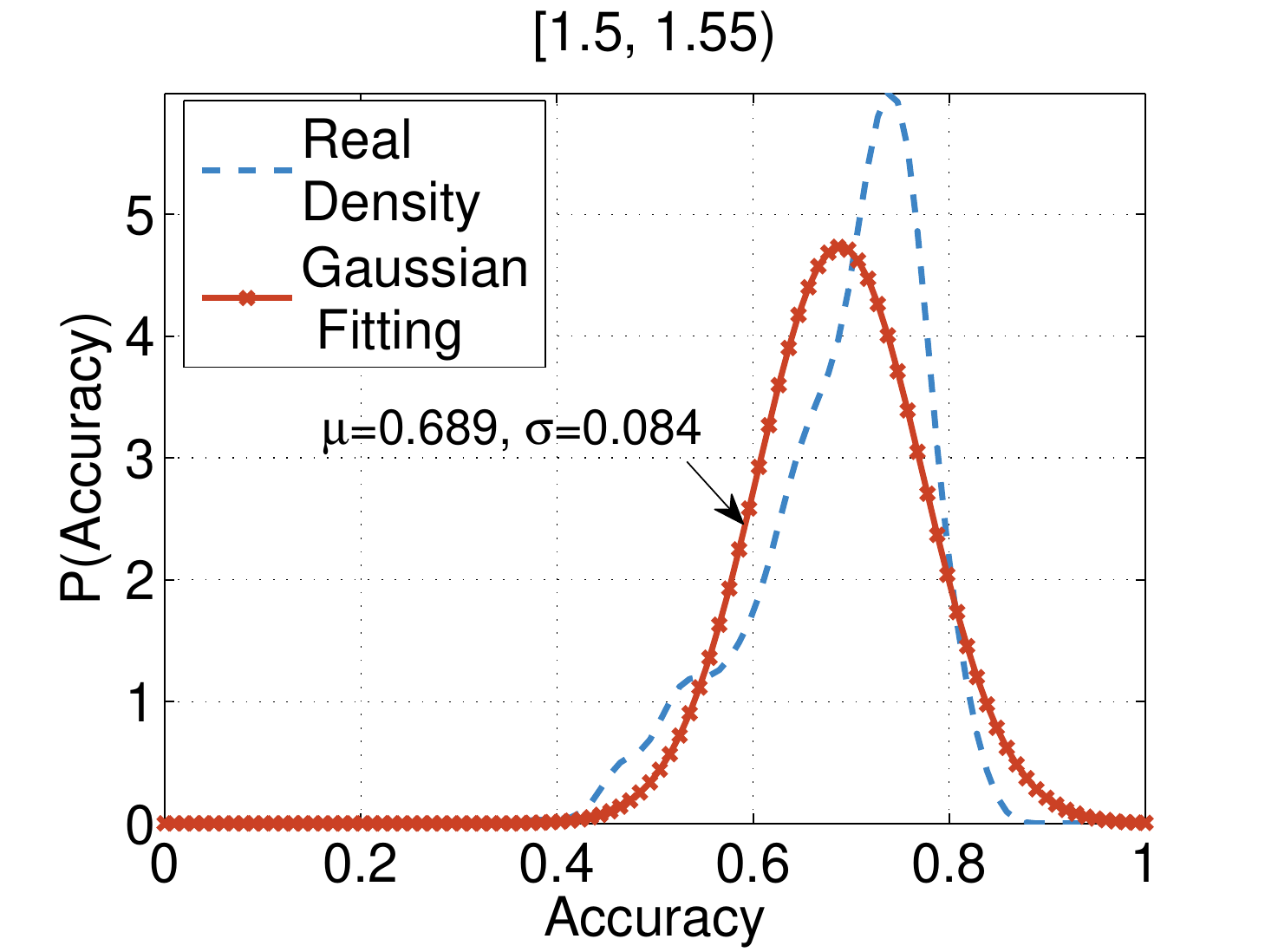}
\end{minipage}
}
\subfigure[]{
\begin{minipage}[b]{0.3\textwidth}
\includegraphics[width=1\textwidth]{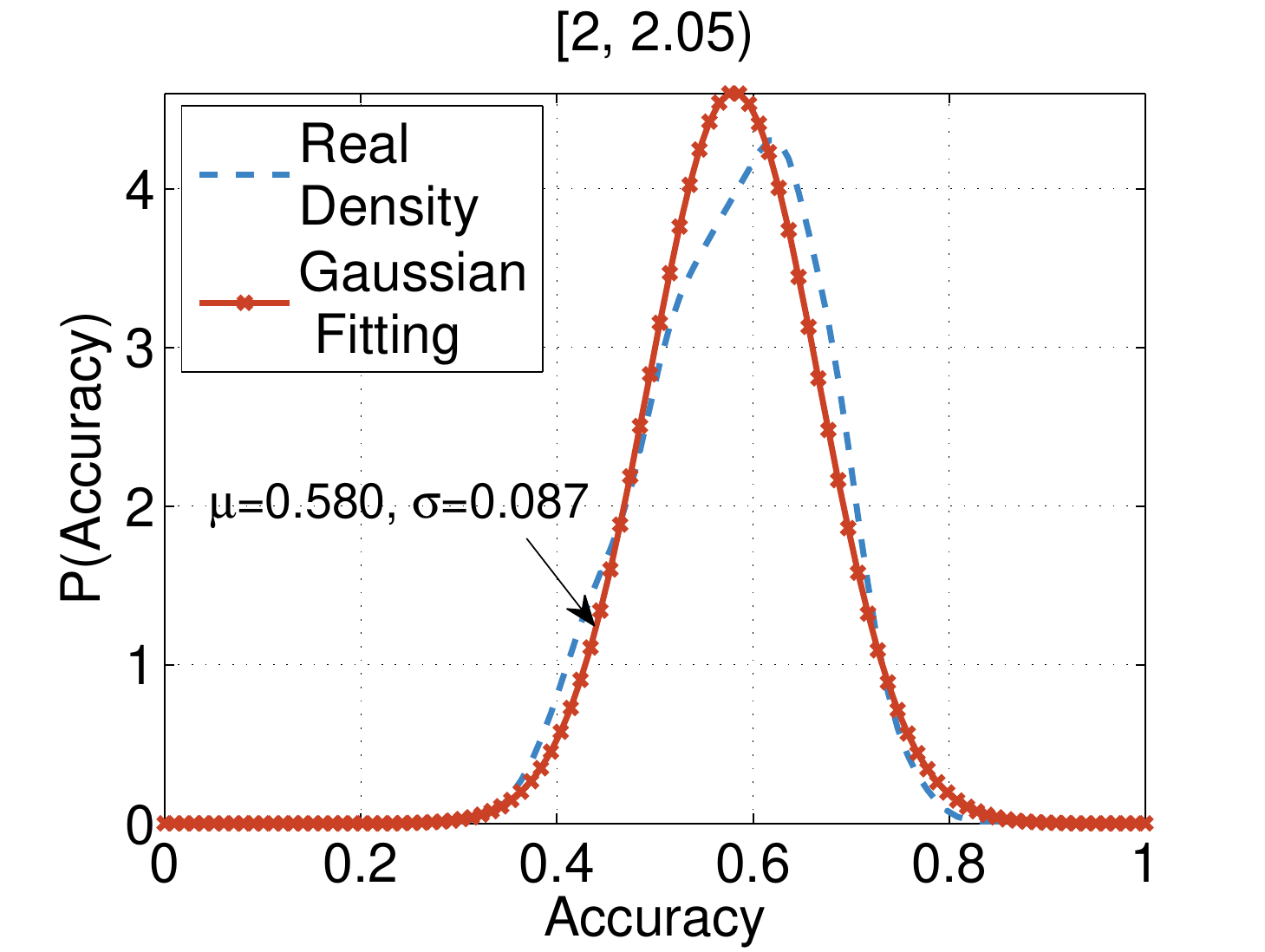}
\end{minipage}
}
\subfigure[]{
\begin{minipage}[b]{0.3\textwidth}
\includegraphics[width=1\textwidth]{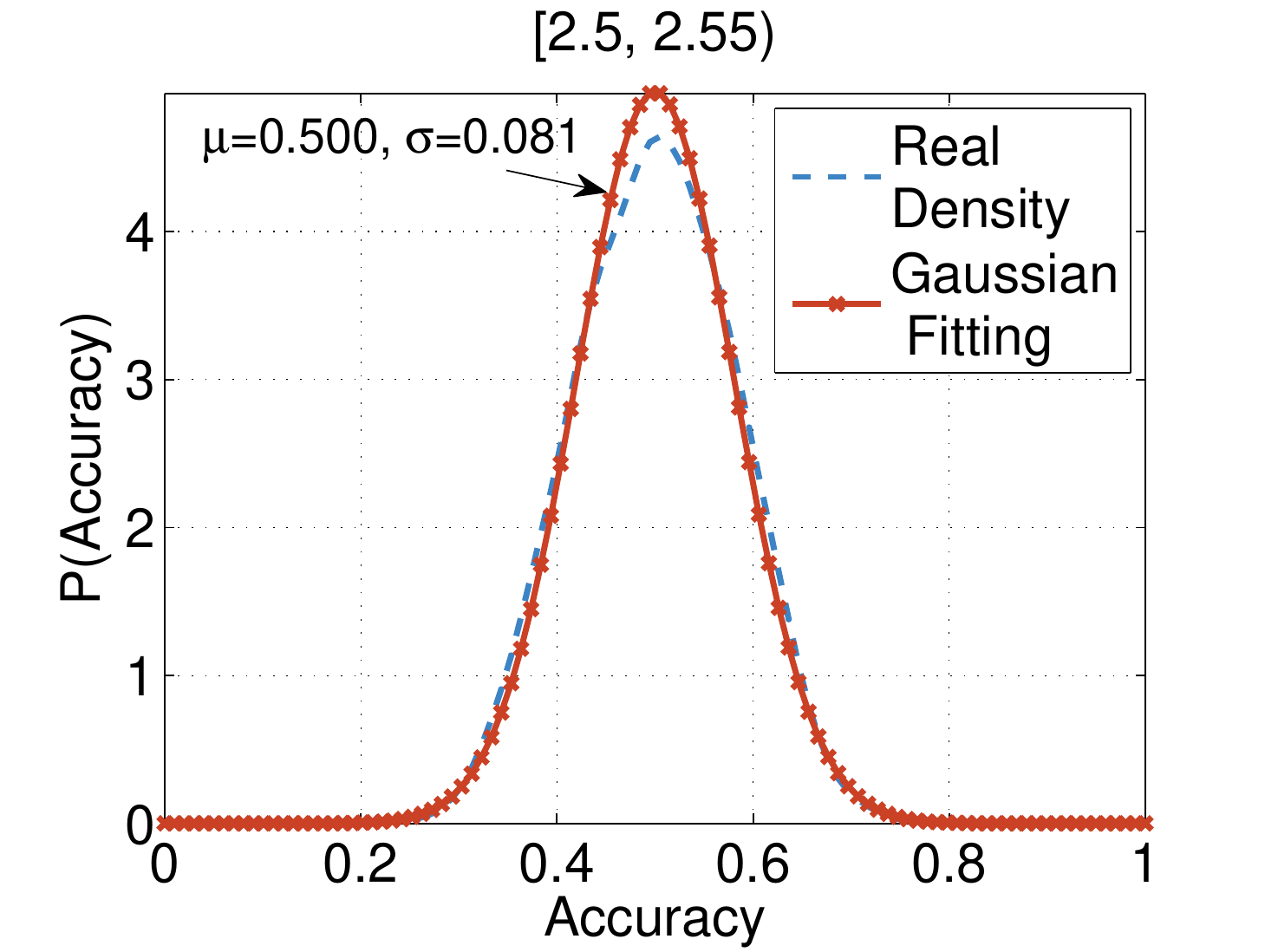}
\end{minipage}
}
\subfigure[]{
\begin{minipage}[b]{0.3\textwidth}
\includegraphics[width=1\textwidth]{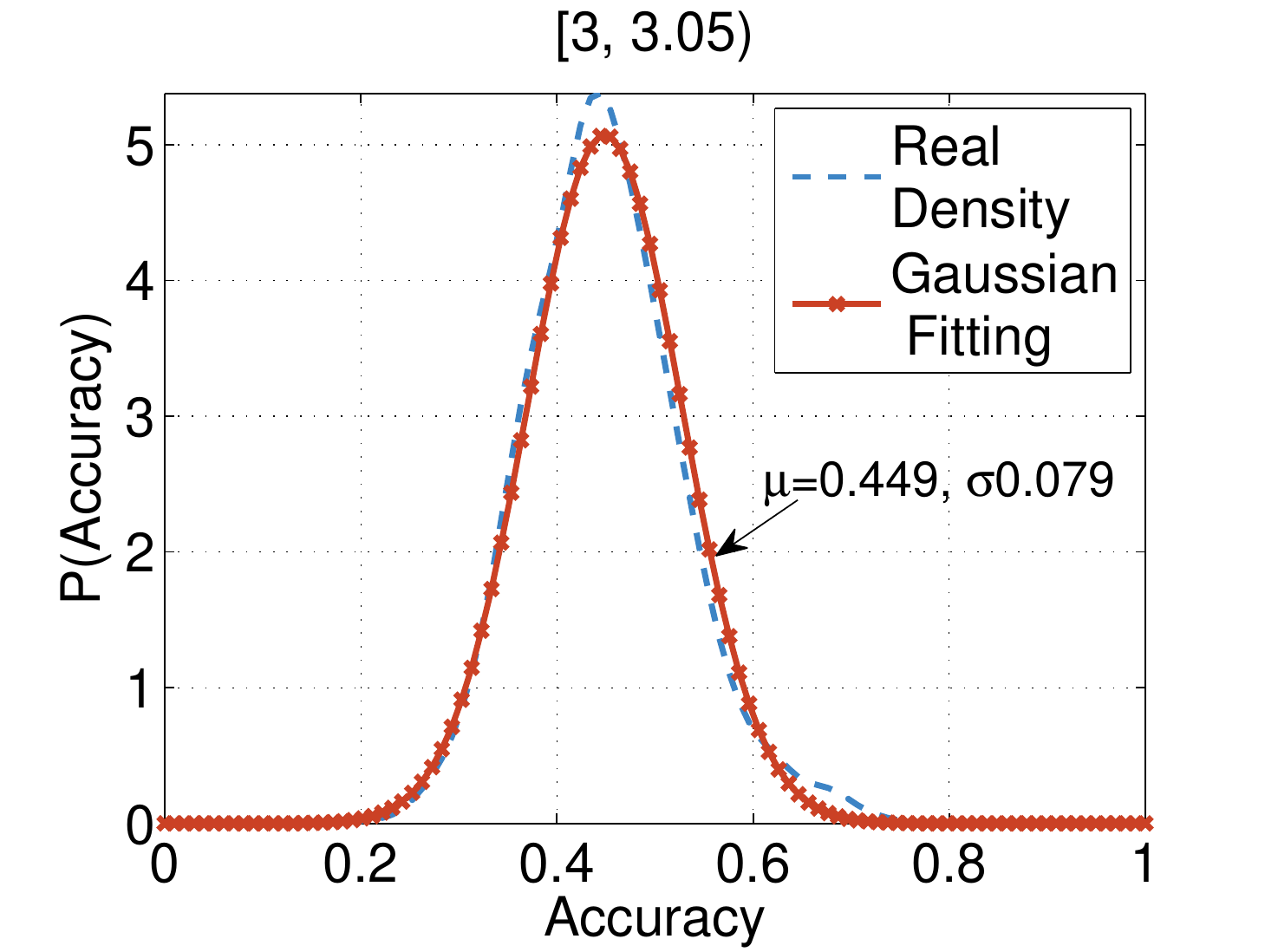}
\end{minipage}
}
\subfigure[]{
\begin{minipage}[b]{0.3\textwidth}
\includegraphics[width=1\textwidth]{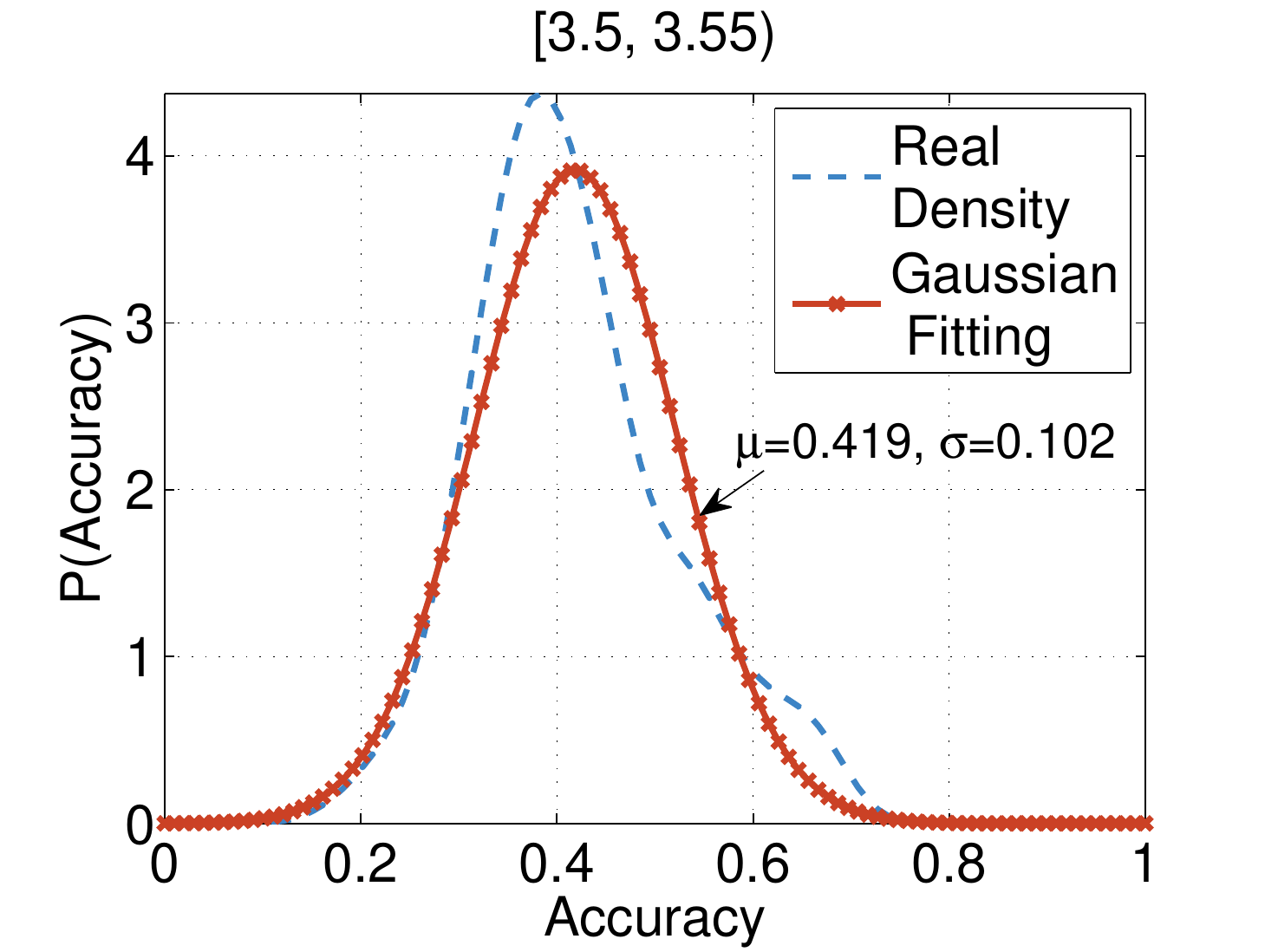}
\end{minipage}
}
\subfigure[]{
\begin{minipage}[b]{0.3\textwidth}
\includegraphics[width=1\textwidth]{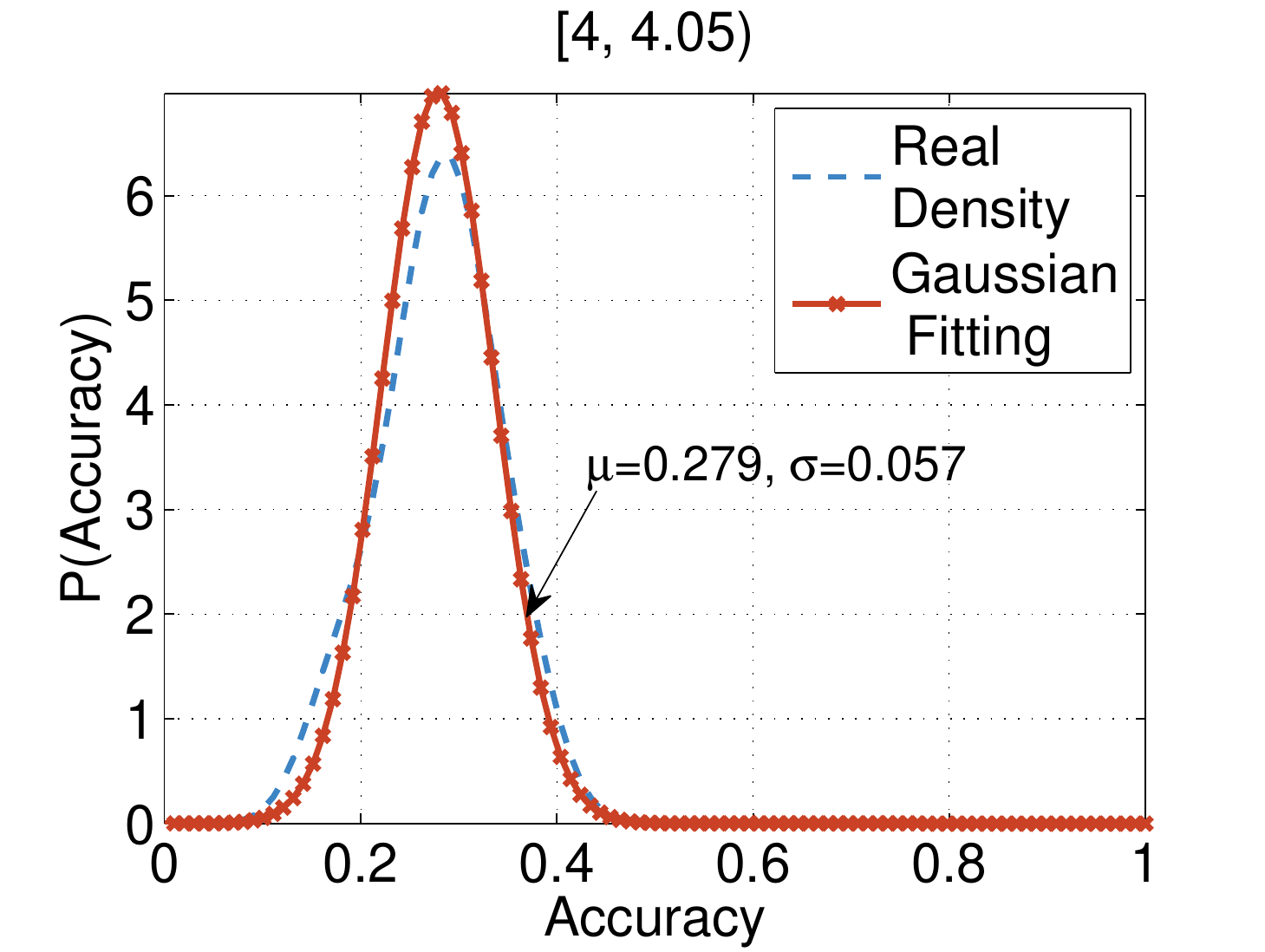}
\end{minipage}
}
\caption{Fitting the probability density distribution of prediction accuracy in entropy intervals with Gaussian distribution. The horizontal axis presents the accuracy, and the vertical axis indicates the real density of the accuracy. (a) Entropy interval [0, 0.05), (b) Entropy interval [0.5, 0.55), (c) Entropy interval [1, 1.05), (d) Entropy interval [1.5, 1.55), (e) Entropy interval [2, 2.05), and (f) Entropy interval [2.5, 2.55), (g) Entropy interval [3, 3.05), (h) Entropy interval [3.5, 3.55), and (i) Entropy interval [4, 4.05).}
\label{Accuracy-fit}
\end{figure*}

We analyze the probability density distribution of accuracy in each entropy interval which is estimated through kernel density estimation and use Gaussian distribution to fit them and obtain the parameters. The parameters for the 84 intervals are calculated, including mean $\mu$ and standard deviation $\sigma$. In order to showing the performance, we take results of 9 intervals as examples shown in Fig. \ref{Accuracy-fit}. For interval [0, 0.05), $\mu=0.999$ and $\sigma=0.002$. For interval [0.5, 0.55), $\mu=0.928$ and $\sigma=0.019$. For interval [1, 1.05), $\mu=0.814$ and $\sigma=0.070$. For interval [1.5, 1.55), $\mu=0.689$ and $\sigma=0.084$. For interval [2, 2.05), $\mu=0.580$ and $\sigma=0.0865$.  For interval [2.5, 2.55), $\mu=0.500$ and $\sigma=0.081$. For interval [3, 3.05), $\mu=0.449$ and $\sigma=0.079$. For interval [3.5, 3.55), $\mu=0.419$ and $\sigma=0.102$. For interval [4, 4.05), $\mu=0.279$ and $\sigma=0.057$. Complete results of parameters are shown as the "$\mu$ of Gaussian Fitting" and "$\sigma$ of Gaussian Fitting" curves in Fig. \ref{Mu-fit} and Fig. \ref{Sigma-fit}. We can observe that $\mu$ decreases with the increasing entropy. However, $\sigma$ shows non-linear relationship with entropy.

\subsection{Least Squares Parameter Estimation}
When we get the parameters of distribution in each entropy interval, mean $\mu$ and standard deviation $\sigma$, which both vary with increasing entropy, it needs to find the correlation between parameters and entropy. For mean value, it shows linear relationship with entropy. However, it is not so simple for standard deviation. So, we try different distributions for standard deviation and make selection according to MSE.

\begin{figure}[!htb]	
\centerline{\includegraphics[width=0.42\textwidth]{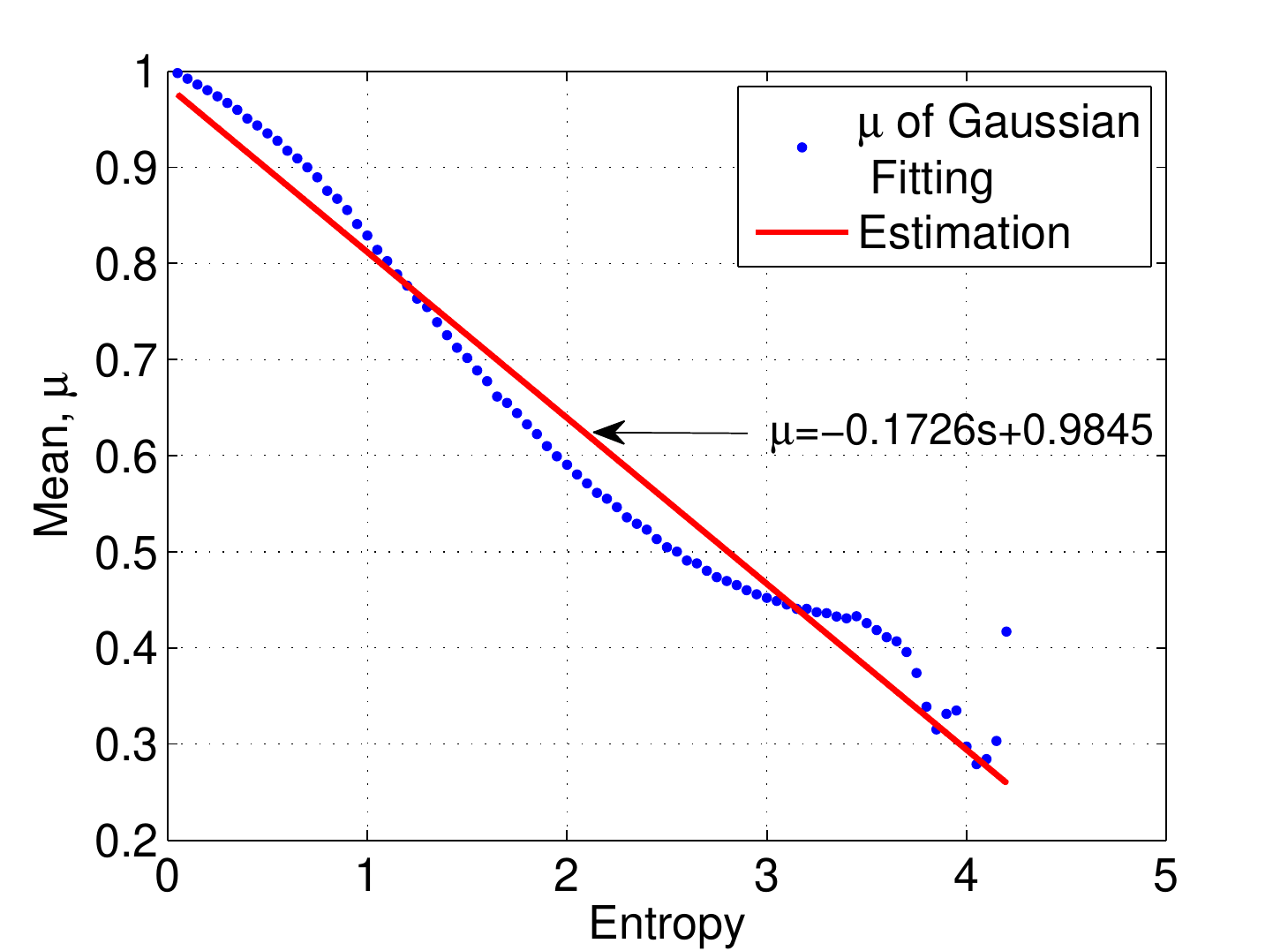}}
\caption{Mean value of Gaussian fitting results and estimation of the mean of different entropy intervals.}
\label{Mu-fit}
\end{figure}

The mean, which can seem as the average of prediction accuracy, decreases with the increasing entropy. The relationship between mean and entropy is similar to linear relationship as shown in Fig. \ref{Mu-fit}. We set the linear function as the basis function. Through least squares method, the mean corresponding to entropy intervals can be described as $\mu=-0.1726s+0.9845$, where $s$ is the upper bound of entropy intervals. The result is shown in Fig. \ref{Mu-fit}. The "$\mu$ of Gaussian Fitting" is the mean of Gaussian fitting result in different entropy intervals. The "Linear" is the estimation result given entropy.

\begin{figure}[!htb]	
\centerline{\includegraphics[width=0.4\textwidth]{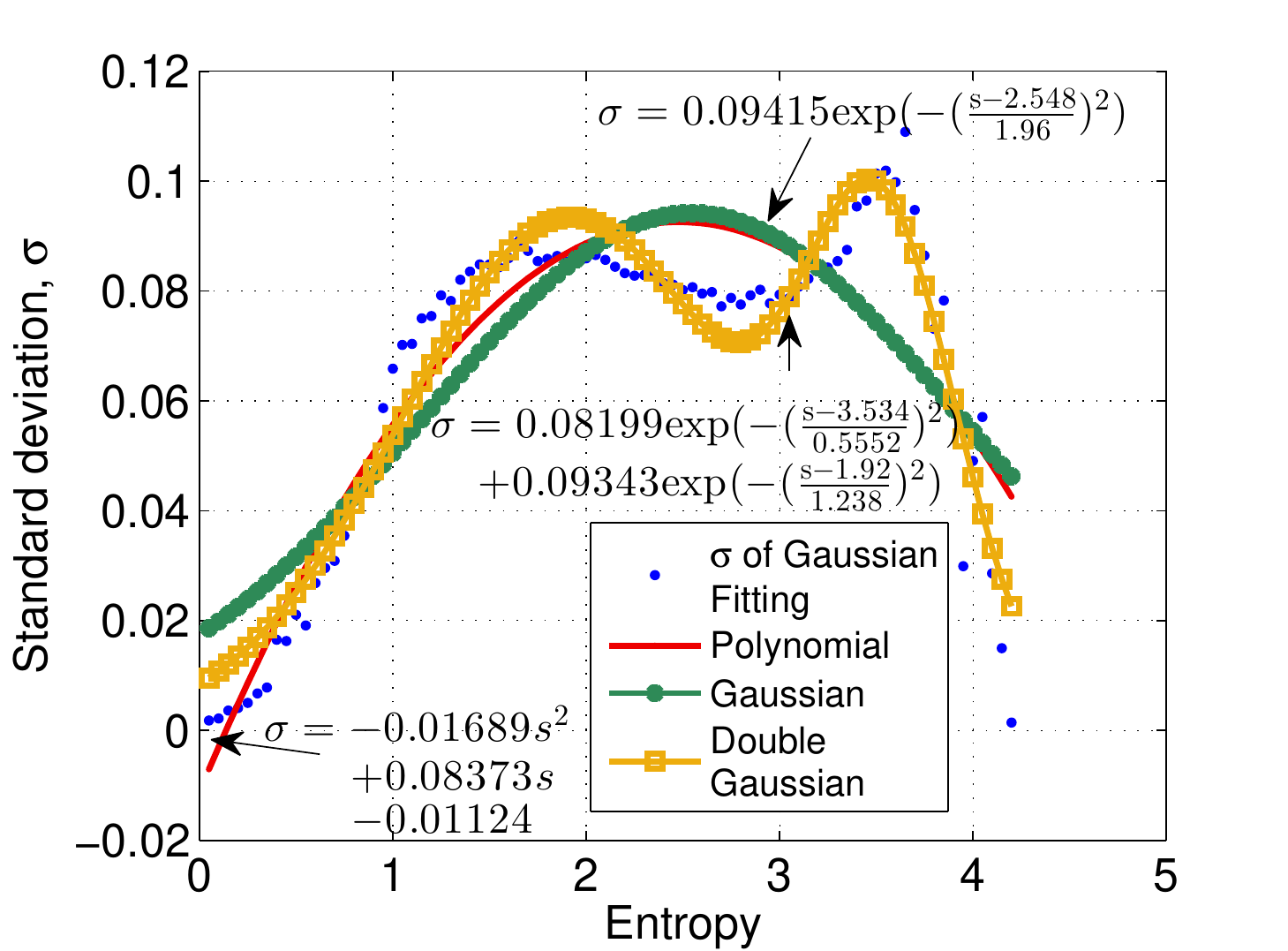}}
\caption{Standard deviation value of Gaussian fitting result and estimation of the standard deviation of different entropy intervals.}
\label{Sigma-fit}
\end{figure}

Contrary to the mean value, standard deviation has up and down with the increasing entropy. Consequently, we have tried different basis functions, including polynomial, Gaussian distribution and double Gaussian distribution. Using least squares method, the standard deviation of entropy intervals can be described as $\sigma=-0.01689s^2+0.08373s-0.01124$, $\sigma=0.09415exp(-(\frac{s-2.548}{1.96})^2)$, $\sigma=0.08199exp(-(\frac{s-3.534}{0.5552})^2)+0.09343exp(-(\frac{s-1.92}{1.238})^2)$, corresponding to "Polynomial", "Gaussian", and "Double Gaussian" in Fig. \ref{Sigma-fit}. The "$\sigma$ of Gaussian Fitting" is the standard deviation of Gaussian fitting result in different entropy intervals.

\begin{figure*}[!htb]
\centering
\subfigure[]{
\begin{minipage}[b]{0.3\textwidth}
\includegraphics[width=1\textwidth]{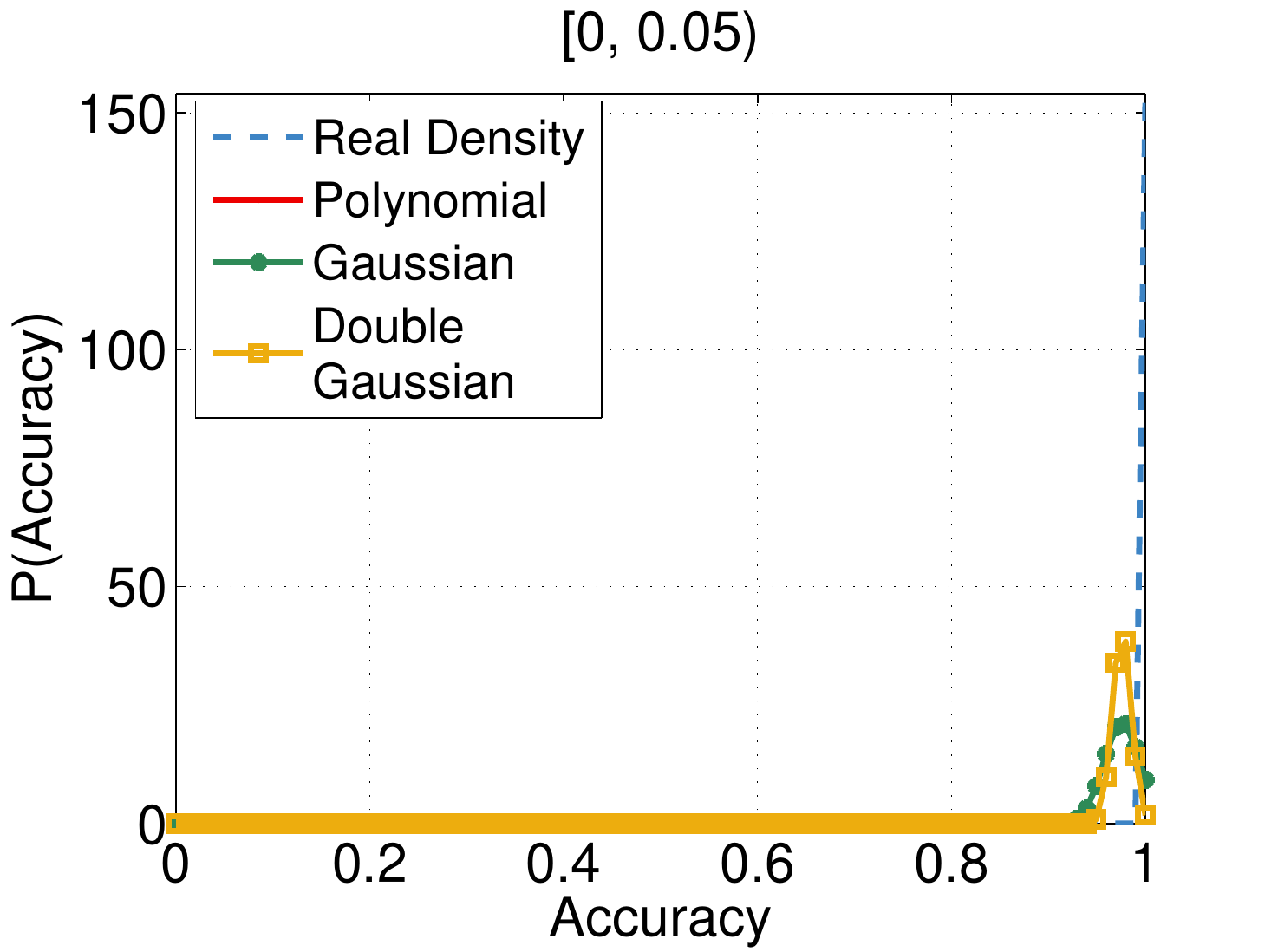}
\end{minipage}
}
\subfigure[]{
\begin{minipage}[b]{0.3\textwidth}
\includegraphics[width=1\textwidth]{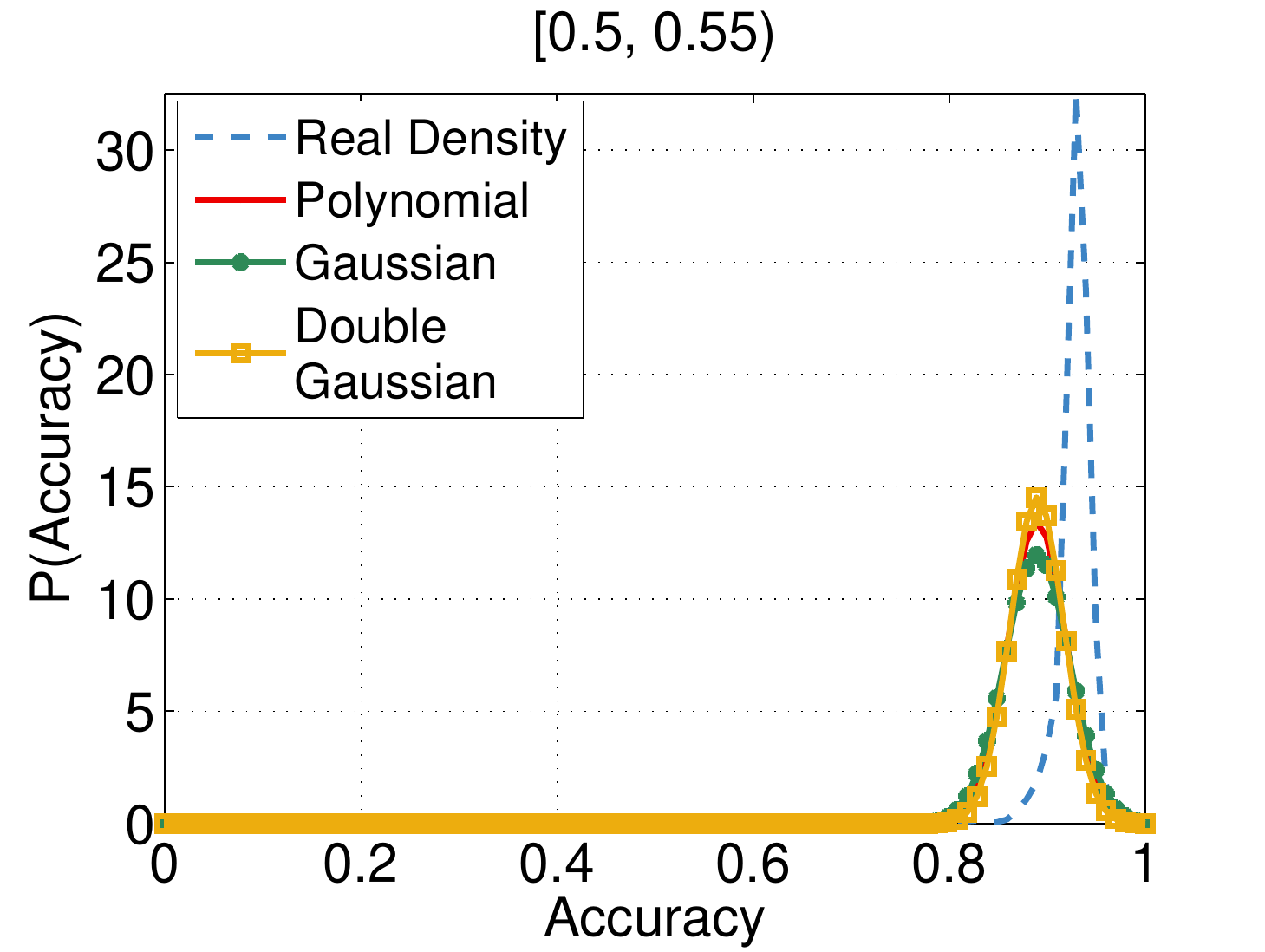}
\end{minipage}
}
\subfigure[]{
\begin{minipage}[b]{0.3\textwidth}
\includegraphics[width=1\textwidth]{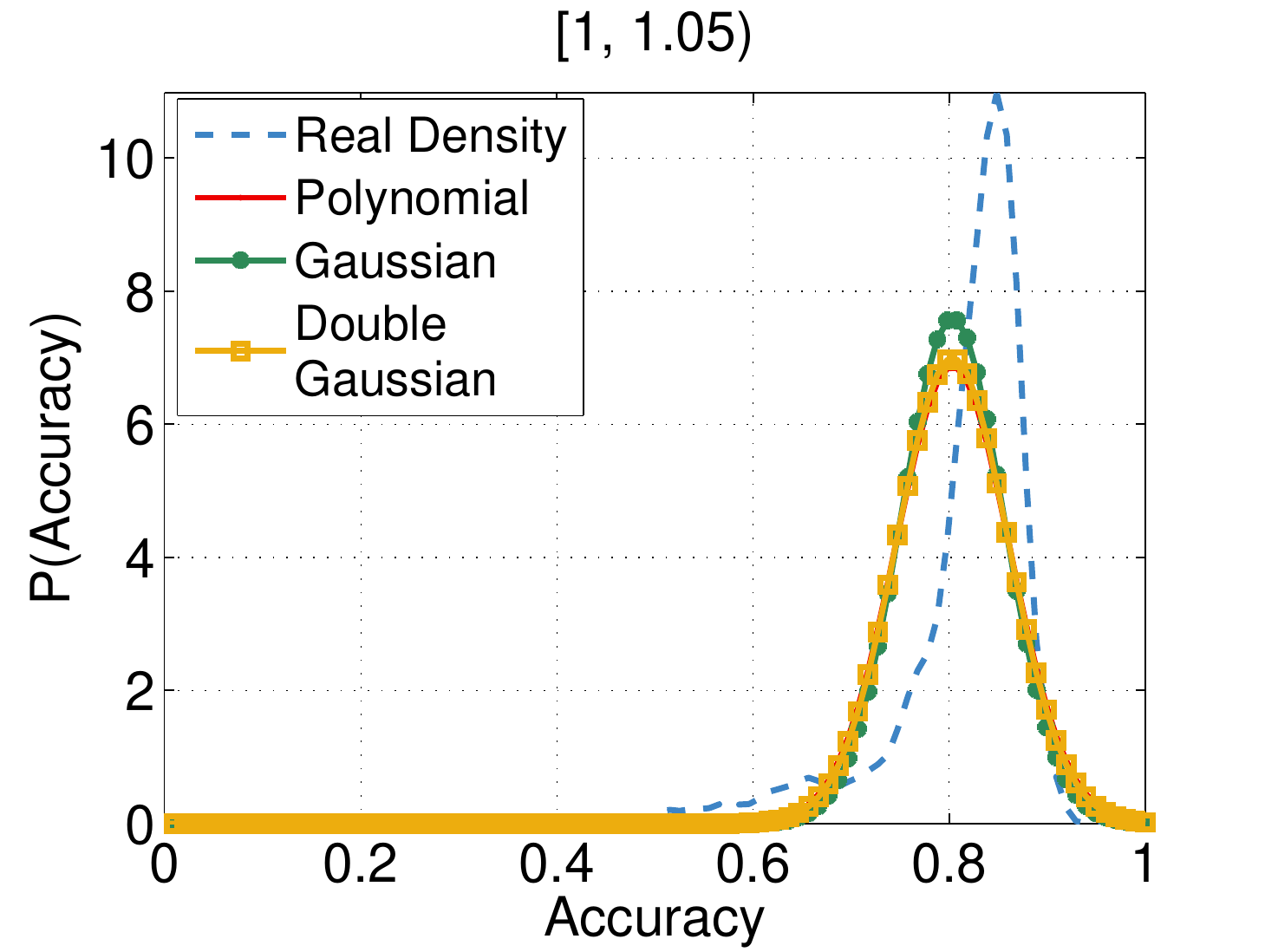}
\end{minipage}
}
\subfigure[]{
\begin{minipage}[b]{0.3\textwidth}
\includegraphics[width=1\textwidth]{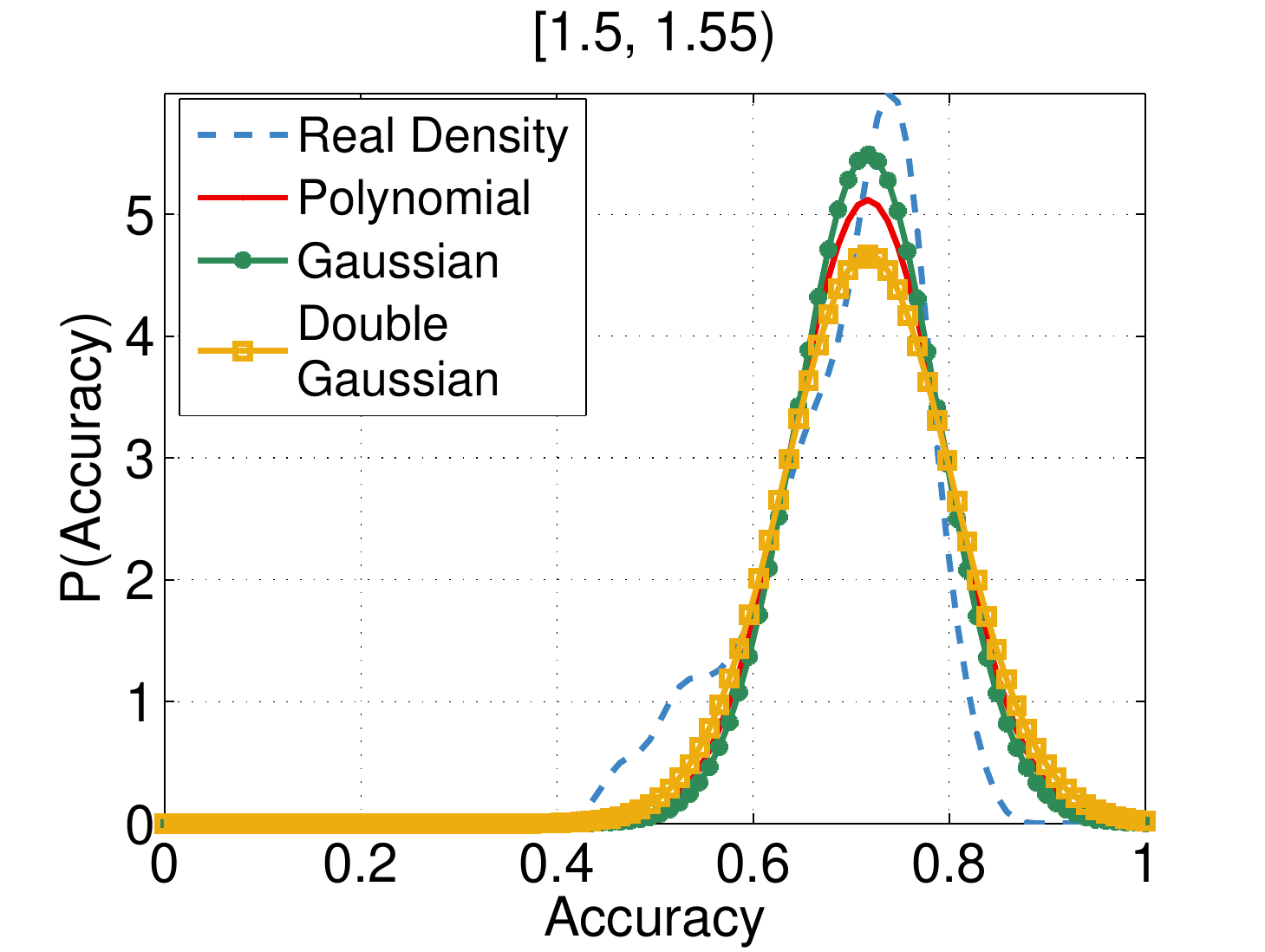}
\end{minipage}
}
\subfigure[]{
\begin{minipage}[b]{0.3\textwidth}
\includegraphics[width=1\textwidth]{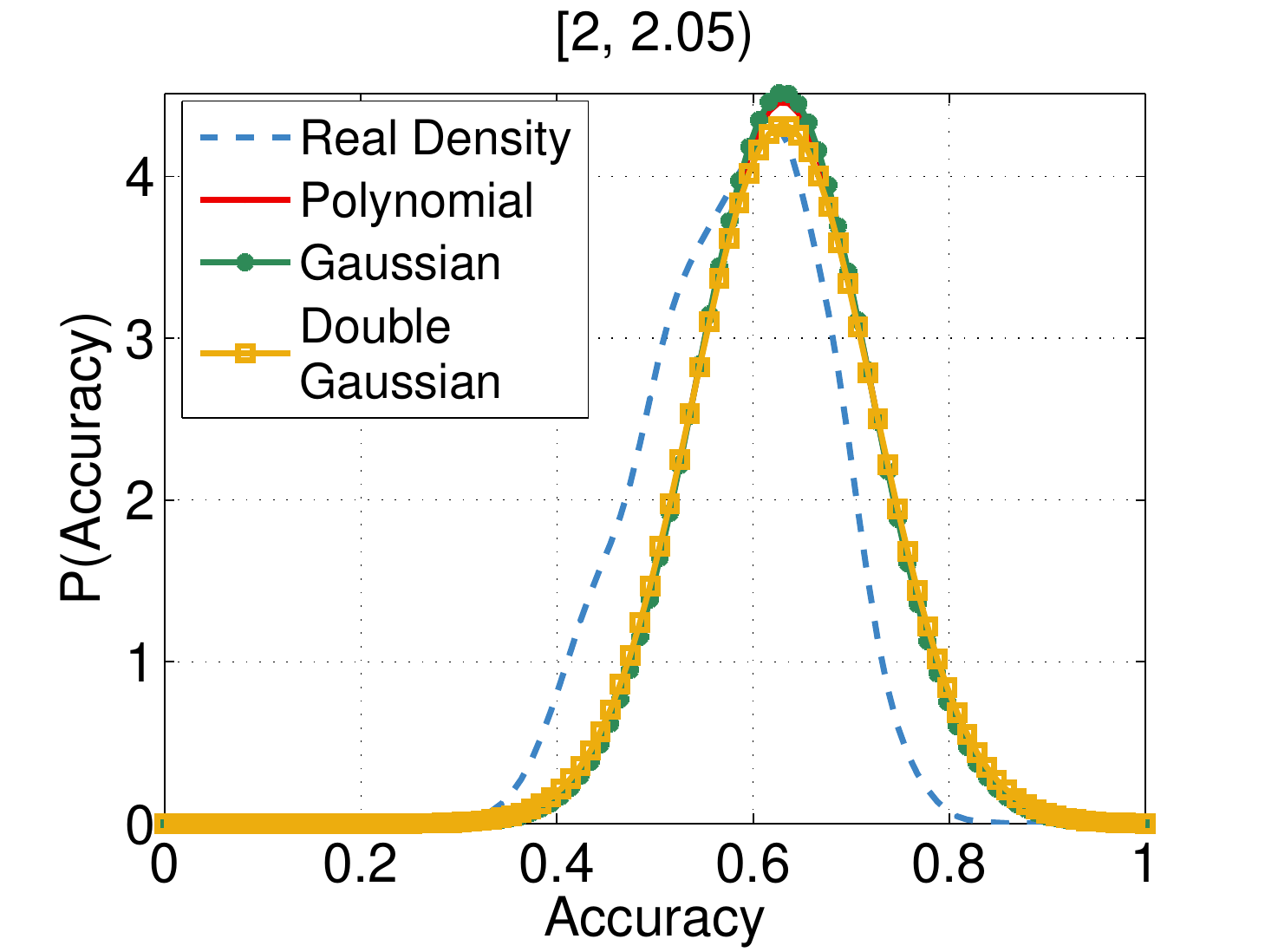}
\end{minipage}
}
\subfigure[]{
\begin{minipage}[b]{0.3\textwidth}
\includegraphics[width=1\textwidth]{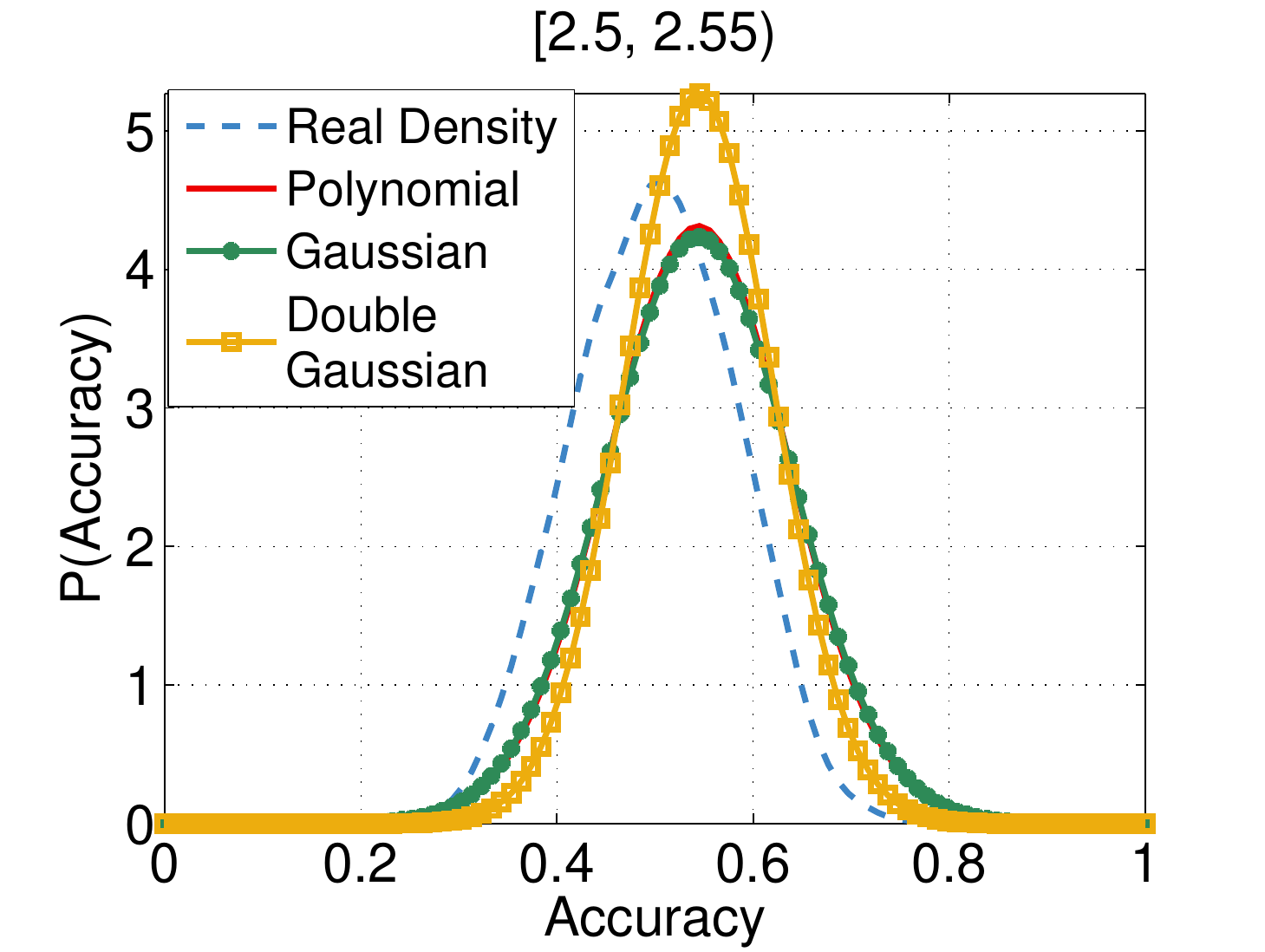}
\end{minipage}
}
\subfigure[]{
\begin{minipage}[b]{0.3\textwidth}
\includegraphics[width=1\textwidth]{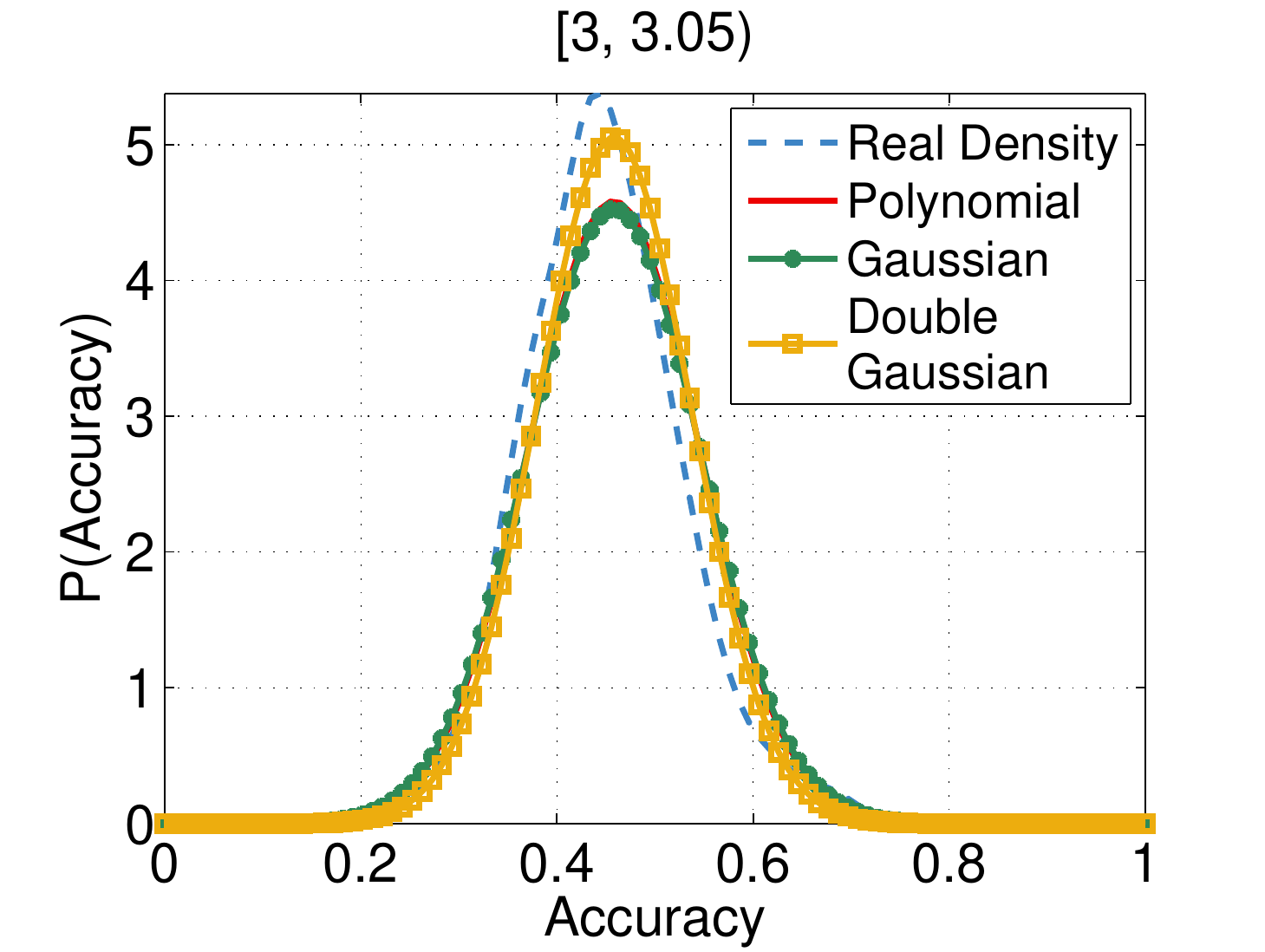}
\end{minipage}
}
\subfigure[]{
\begin{minipage}[b]{0.3\textwidth}
\includegraphics[width=1\textwidth]{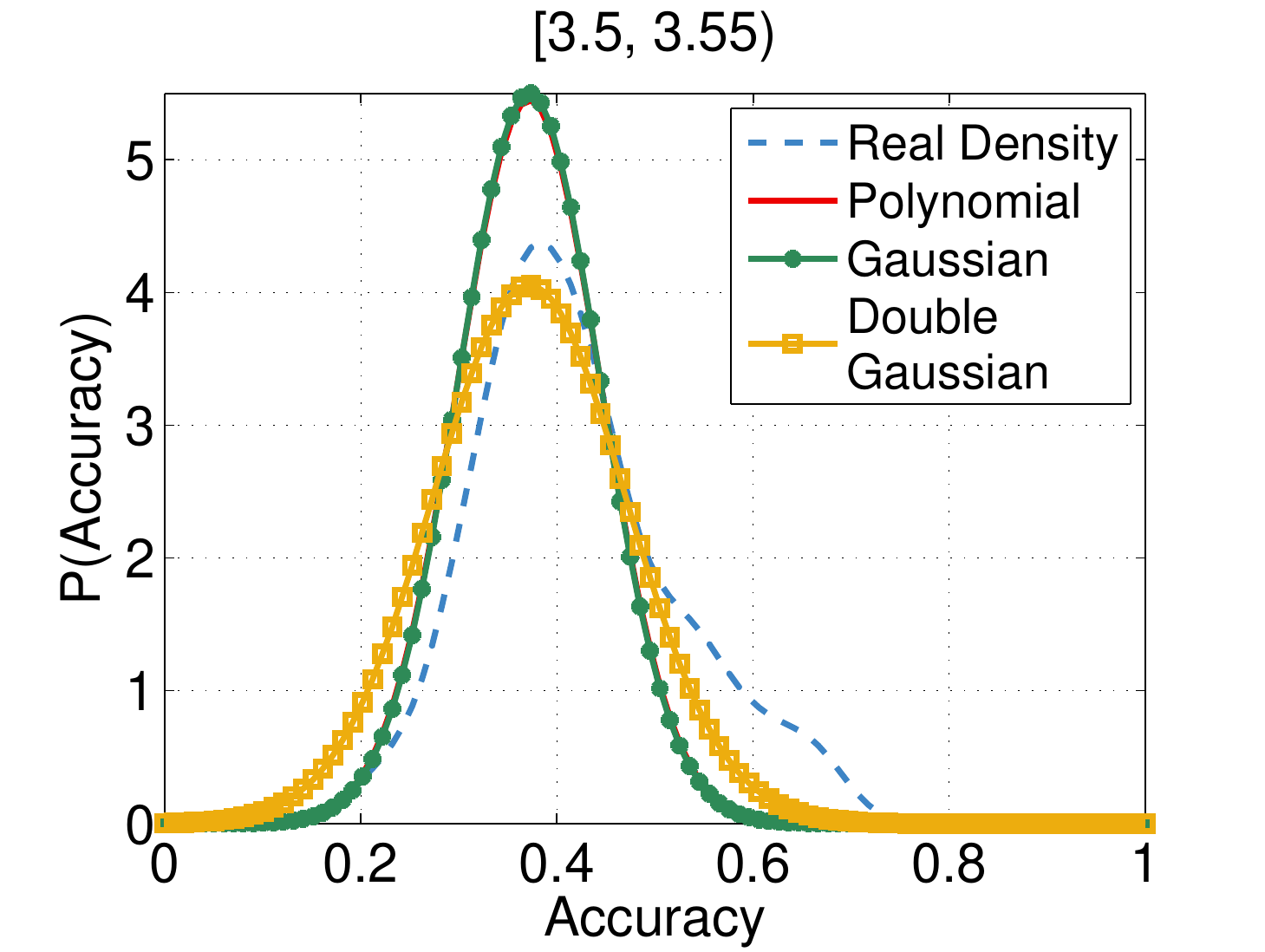}
\end{minipage}
}
\subfigure[]{
\begin{minipage}[b]{0.3\textwidth}
\includegraphics[width=1\textwidth]{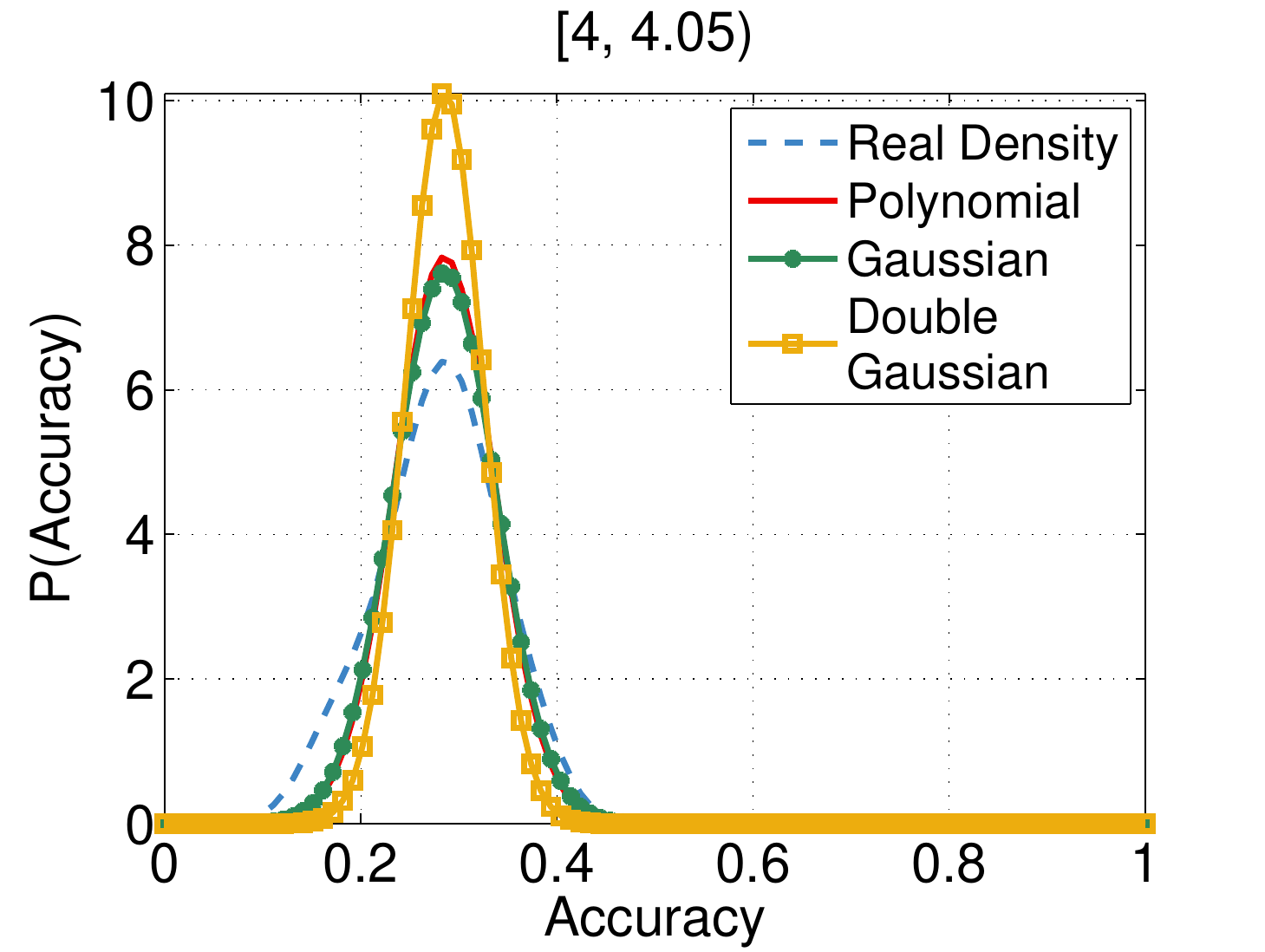}
\end{minipage}
}
\caption{Based on the parameter estimation, the estimation of probability density distribution of accuracy in entropy intervals. (a) Entropy interval [0, 0.05), (b) Entropy interval [0.5, 0.55), (c) Entropy interval [1, 1.05), (d) Entropy interval [1.5, 1.55), (e) Entropy interval [2, 2.05), and (f) Entropy interval [2.5, 2.55), (g) Entropy interval [3, 3.05), (h) Entropy interval [3.5, 3.55), and (i) Entropy interval [4, 4.05).}
\label{Parameter-fit}
\end{figure*}

Based on parameter estimation results, we can obtain the estimation of probability density function of accuracy corresponding to each entropy interval. We also take the same 9 intervals as examples to show the estimation results in Fig. \ref{Parameter-fit}.

In order to compare the performance of different estimating results for the standard deviation, we calculate the MSE respectively. The three estimations perform well in the middle of the range of entropy but perform a little poor at the both ends of the range. The MSE of the "Polynomial" is the largest, and the MSE of the "Gaussian" is smallest, which says that the "Gaussian" have the best performance which can be used to describe the relationship between standard deviation and entropy.

\begin{figure}	
\centerline{\includegraphics[width=0.4\textwidth]{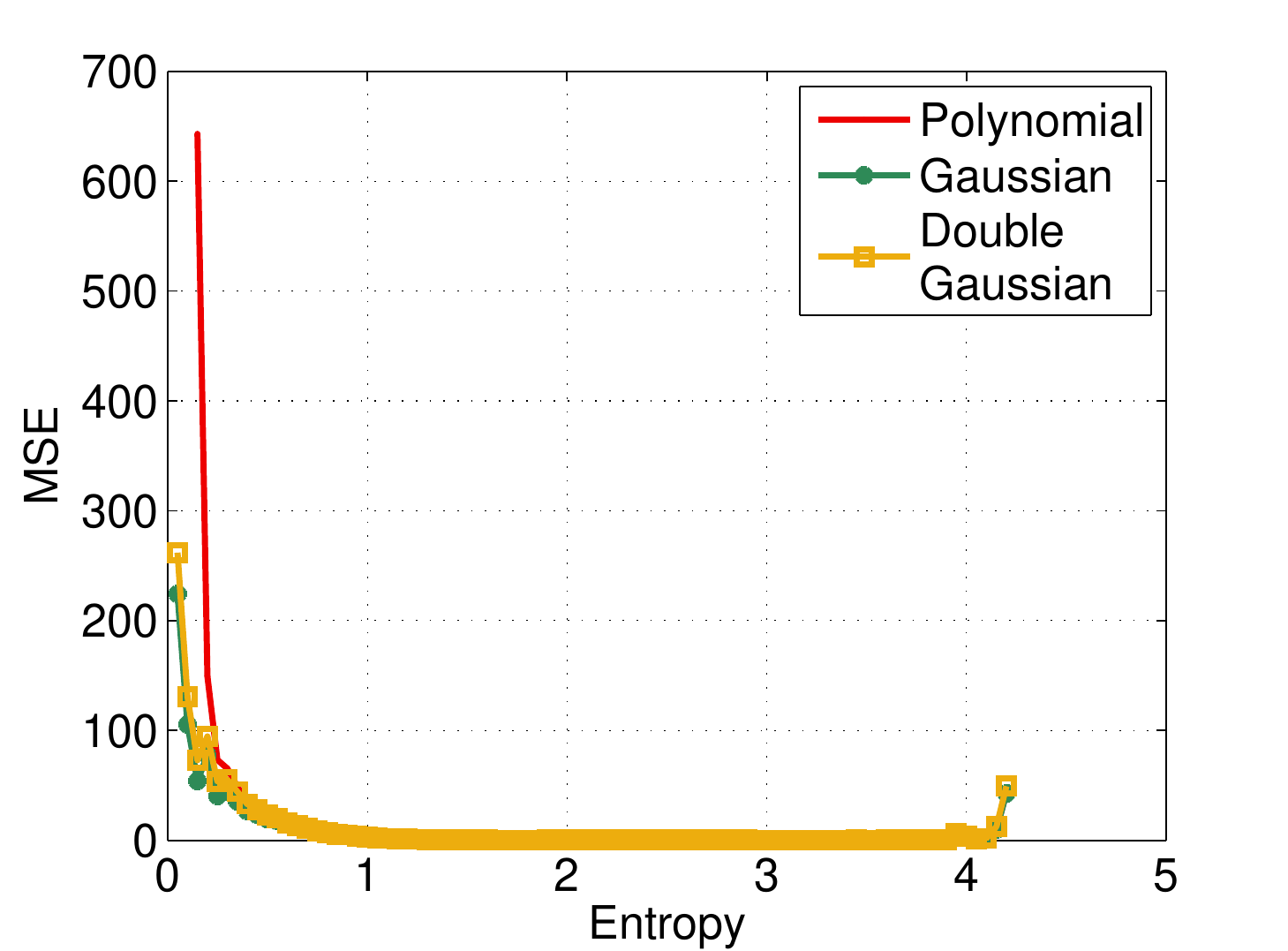}}
\caption{MSE(mean square error) of the three estimation methods.}
\label{MSE}
\end{figure}

We use Kolmogorov-Smirnov test to check whether the Gaussian distribution based on the estimated $\mu$ and $\sigma$ follows the same distribution with the original distribution of accuracy. The testing results demonstrate the distributions based the estimated parameters follows the same distribution with the original distributions of accuracy on majority of the intervals. However, on 30\% entropy intervals the results can't pass the test, which means that the distribution on these intervals can't be represented by the estimated result. It's due to that the distribution of accuracy on these intervals are not completely asymmetrical. The Gaussian distribution based on the estimated results is just an approximation of the original distributions.

\subsection{Drawing A Functional Gaussian Distribution}
According to the parameters estimation, we obtain the parameters given entropy. Based on the expression form of Gaussian distribution, we can draw a functional Gaussian distribution to describe the probability density function of accuracy given the entropy interval in Eq. \ref{acc}.

\begin{equation}
\begin{aligned}
\label{acc}
&p(x|s)=\frac{1}{\sqrt{2\pi}\sigma(s)}exp(-\frac{(x-\mu(s))^2}{2(\sigma(s))^2})\\
&\mu(s)=-0.1726s+0.9845\\
&\sigma(s)=0.09415exp(-(\frac{s-2.548}{1.96})^2)\\
&0\leq x\leq 1, s=\{0.05n|n=1,2,...,84\}.
\end{aligned}
\end{equation}
where $x$ is prediction accuracy and $s$ is predictability entropy. $x$ is a continuous variable, while $s$ is a discrete variable as the analysis is based on entropy intervals. Given a entropy value $s$, we can calculate the mean $\mu$ and standard deviation $\sigma$, based on which it is easy to get probability density function of accuracy. Using the probability density function, the range and the probability distribution of accuracy can be calculated.

\section{Discussions}
\label{sec5}
First, our work focus on the correlation between accuracy and entropy, and we find a formula to describe the correlation based on a real data set and a certain prediction model. Given an entropy interval, we can obtain the probability density function of accuracy that can be achieved. It is different from previous work\cite{predictability}, which deduces the upper bound and lower bound of accuracy based on entropy.

Second, we propose an approach to analyze the correlation between achieved accuracy and entropy. Although our work is not general enough, it may shed light on others. They can verify it with another data sets and another prediction models. The correlation between the parameters and entropy may appear differently. However, our work provide a method to study prediction accuracy.

\section{Conclusions}
\label{sec6}
In this work, we model the accuracy based on entropy. As the users with the same level of entropy achieve different accuracies, we divide the entropy into intervals and analyze the probability density distribution of accuracy in each interval. All of the distributions follow unimodel distributions. We fit them with Gaussian distribution and obtain the mean and standard deviation of each interval. Then, we find the correlation between the parameters and entropy intervals. The mean can be modelled as $\mu=-0.1726s+0.9845$, and standard deviation can be modelled as $\sigma=0.09415exp(-(\frac{s-2.548}{1.96})^2)$, where $s$ is the discretized entropy. Based on the correlation, we draw a functional Gaussian distribution to describe the probability density function of accuracy given entropy.

Although we propose a formula to describe the correlation based on a real data set and a certain prediction model, it is not accurate and general enough. In the future work, we will try to find more accurate expression of the accuracy distribution and focus on more general conditions and deduce more general results.

\section*{acknowledgment}
This work was partially supported by Key Program of Natural Science Foundation of China under Grant(61631018), the Fundamental Research Funds for the Central Universities, Huawei Technology Innovative Research.

%\section*{acknowledgment}
%This work was partially supported by Key Program of Natural Science Foundation of China under Grant(61631018),, the Fundamental Research Funds for the Central Universities(WK3500000003), Huawei Technology Innovative Research (YB2013120167).

%\bibliographystyle{IEEEtran}
%\bibliography{123}
\balance

\end{document}